\documentclass[twocolumn,apj,numberedappendix,appendixfloats,twocolappendix]{openjournal}

\usepackage{newtxtext,newtxmath,amsmath}
\usepackage[T1]{fontenc}
\usepackage{graphicx}
\usepackage{enumitem}
\usepackage{xcolor}
\usepackage{hyperref}
\usepackage{gensymb}
\usepackage{CJK}
\usepackage{tabularx}
\usepackage{orcidlink}
\hypersetup{
    unicode, 
    colorlinks=true,
    linkcolor=linkcolor,
    citecolor=linkcolor,
    filecolor=linkcolor,
    urlcolor=linkcolor,
}
\definecolor{linkcolor}{rgb}{0.0,0.3,0.5}
\usepackage{nameref}
\usepackage{color,colortbl}
\definecolor{linkcolor}{rgb}{0.0,0.3,0.5}
\usepackage{tensind}
\tensordelimiter{?}
\DeclareGraphicsExtensions{.bmp,.png,.jpg,.pdf}
\usepackage{verbatim}
\usepackage[normalem]{ulem}
\usepackage{soul}
\usepackage[caption=false]{subfig}

\newcolumntype{C}{>{\centering\arraybackslash}X}

\graphicspath{ {./figs/} }

\begin{document}

\title{Detecting Strongly-Lensed Supernovae in Wide-field Space Telescope Imaging via Deep Learning}

\author{\vspace{-1.6cm}Fawad Kirmani$^{1,*}$}
\author{Arjun Karki$^{2}$}
\author{Steve Rodney$^{2}$}
\author{Kyle Lackey$^{2}$}
\author{Varsha P. Kulkarni$^{2}$}
\author{John R. Rose$^{1}$}
\author{Justin Pierel$^{3}$}

\affiliation{$^{1}$Department of Computer Science and Engineering, University of South Carolina, Columbia, SC, USA}
\affiliation{$^{2}$Department of Physics and Astronomy, University of South Carolina, Columbia, SC, USA}
\affiliation{$^{3}$Space Telescope Science Institute, Baltimore, MD, USA}

\thanks{E-mail: \href{mailto:fkirmani@email.sc.edu}{fkirmani@email.sc.edu}}
\thanks{E-mail: \href{mailto:karkia@email.sc.edu}{karkia@email.sc.edu}}
\thanks{E-mail: \href{mailto:steve.rodney@gmail.com}{steve.rodney@gmail.com}}
\thanks{E-mail: \href{mailto:kelackey@gmail.com}{kelackey@gmail.com}}
\thanks{E-mail: \href{mailto:kulkarni@sc.edu}{kulkarni@sc.edu}}
\thanks{E-mail: \href{mailto:rose@cse.sc.edu}{rose@cse.sc.edu}}
\thanks{E-mail: \href{mailto:jpierel@stsci.edu}{jpierel@stsci.edu}}

%%%%%%%%%%%%%%%%%%%%%%%%%%%%%%%%%%%%%%%%%%%%%%%%%%%%%%%%%%%%%%%%%%%%%%%%%%%%%%%%
\begin{abstract}

Gravitationally lensed supernovae (SNe) are extremely rare and fade quickly; as a result, they are challenging to detect. To identify lensed SNe in large imaging datasets, current surveys primarily rely on the {\it magnification} effect of gravitational lensing---searching for transients that appear brighter than expected \cite{c3}. In this work, we present a proof-of-concept study that uses a deep neural network to classify previously detected transients. Instead of relying on magnification, this network aims to identify doubly-imaged SNe with small separations ($<0.6$ arcsec) based on the {\it distorted shape} of the transient object. This proposed method is most applicable to space-based imaging surveys from wide-field imaging observatories such as the upcoming Roman Space Telescope. To train and test our network, we use archival Hubble Space Telescope (HST) imaging surveys. Due to the extreme rarity of lensed SNe, we cannot train a neural network on actual lensed SN data. Instead, we have used HST imaging data to generate simulated imaging datasets for both training and testing. Our simulations use astrophysical priors to define the separations, relative brightnesses, and colors of each multiply-imaged SN. We have also simulated false positives (image artifacts and unlensed supernovae), which are much more prevalent than true lensed SN.  Our deep learning model is trained to identify lensed SNe from a single difference image (i.e., not using multiple epochs). This network achieves a recall score of 99\% on simulated gravitationally lensed SNe. The network successfully distinguishes between single supernovae (SNe) and those with gravitationally lensed SNe, as well as images with zero SNe, achieving recall scores of 90\% and 96\% for single-SNe and zero-SNe images, respectively.

\end{abstract}

%%%%%%%%%%%%%%%%%%%%%%%%%%%%%%%%%%%%%%%%%%%%%%%%%%%%%%%%%%%%%%%%%%%%%%%%%%%%%%%%
\section{{INTRODUCTION}}

\begin{figure*}[!]
    \centering
    \subfloat[]
        \centering
        \includegraphics[scale=0.16]{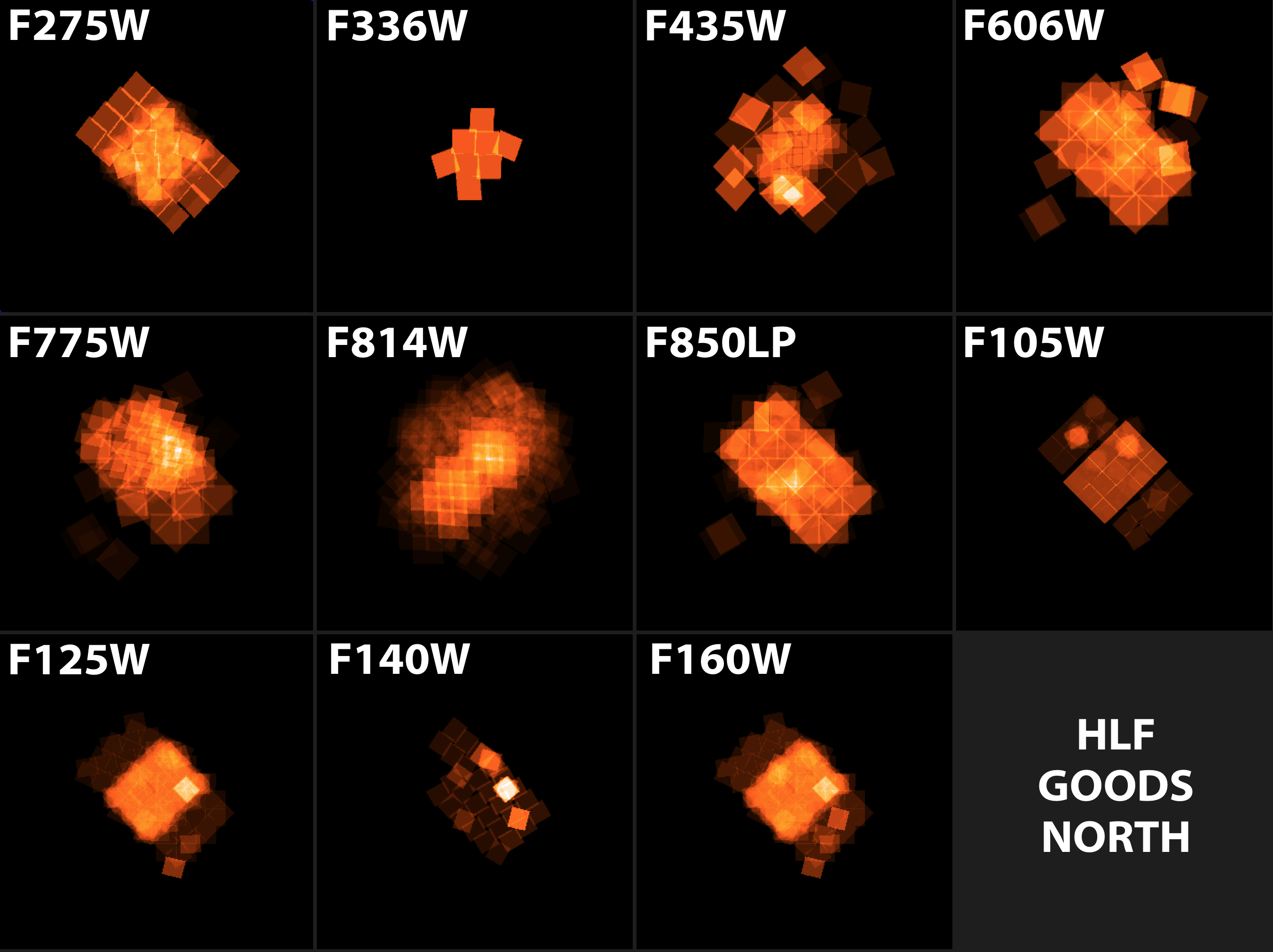}
        \label{fig:hlf_goodsn}
    \subfloat[]
        \centering
        \includegraphics[scale=0.18]{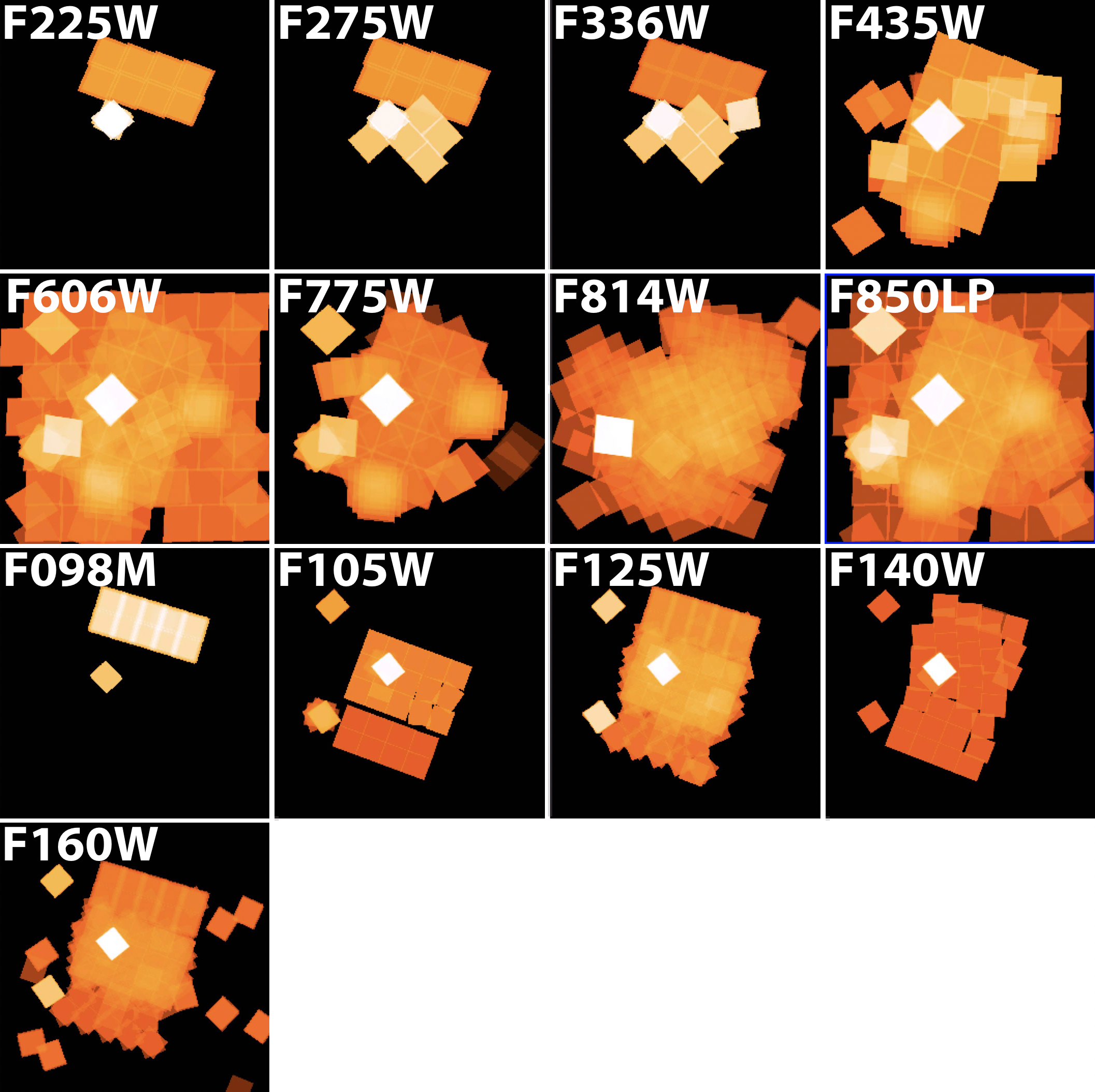}
        \label{fig:hlf_goodss}
    \caption{(a) The exposure time maps of the regions covered by each filter in the HLF-GOODS-N HLSP dataset. The filters include a total of 11 WFC3/UV, ACS/WFC and WFC3/IR filters (F275W, F336W, F435W, F606W, F775W, F814W, F850LP, F105W, F125W, F140W \& F160W). The maps here total ~5800 exposures that comprise the HLF-GOODS-N dataset \cite{c16}. (b) The exposure time maps of the regions covered by each filter in the HLF-GOODS-S HLSP dataset. The XDF/HUDF region is in white, indicating the deepest data. The filters include a total of 13 WFC3/UV, ACS/WFC and WFC3/IR filters (F225W, F275W, F336W, F435W, F606W, F775W, F814W, F850LP, F098M, F105W, F125W, F140W \& F160W). The maps here total ~7500 exposures that comprise the HLF-GOODS-S dataset \cite{c16}.}\label{fig:hlf_goodsn_goodss}
\end{figure*}

Sometimes, the light from a distant supernova (SN) passes very close to an intervening galaxy on its way to Earth. In such cases, that galaxy can act as a gravitational lens, distorting spacetime and acting as an imperfect cosmic magnifying glass. This causes the SN to appear as two or more images in the sky, i.e., a multiply-imaged SN. This event was observed for the first time in 2015 \cite{c4}. 

Lensed SNe fade quickly and must be identified quickly to facilitate timely analysis and spectroscopic follow-up. Due to their extreme rarity, very wide-field and deep imaging surveys are required to detect an appreciable number.  Deep learning methods are, therefore, an attractive tool for automating the detection and classification of lensed SNe from a "big data" stream.  Morgan et al. \cite{c9} designed ZipperNet for the classification of lensed SNe based on their light curves. This is a multi-branch deep neural network that combines convolutional layers (flattened into one-dimensional arrays and condensed in size) with long short-term memory layers (for the lightcurves). ZipperNet was trained on high-fidelity simulations of three ground-based imaging survey data sets: the Dark Energy Survey, Rubin Observatory's Legacy Survey of Space and Time (LSST), and a Dark Energy Spectroscopic Instrument (DESI) imaging survey. ZipperNet was tested on the task of classifying objects from four categories: no lens, galaxy-galaxy lens, lensed Type Ia SN, and lensed core-collapse SN. On an LSST-like dataset, ZipperNet classifies lensed SNe with a ROC-AUC score of 0.97, predicts the spectroscopic type of the lensed SNe with 79\% accuracy, and demonstrates similarly high performance for lensed SNe 1-2 epochs after the first detection.

Putting this concept into practice, Morgan et al. \cite{c10} searched for gravitationally lensed supernovae systems in observed Dark Energy Survey (DES) 5-year SN fields -- 10 3-sq. deg. regions of the sky imaged in the  \textit{griz}  bands approximately every six nights for five years. To perform the search, the authors used the DeepZipper \cite{c9} approach. The authors found that their method achieves a lensed supernova recall of 61.13\% and a false-positive rate of 0.02\% on the DES SN field data. DeepZipper selected 2,245 candidates from a magnitude-limited catalog ( mi \textless 22.5) of 3,459,186 systems. The authors also used human visual inspection to review the systems selected by the network and found three candidate lensed SNe in the DES SN fields.

Tran et al. \cite{c11} presented a spectroscopic confirmation of candidate strong gravitational lenses using the Keck Observatory and Very Large Telescope as part of the ASTRO 3D Galaxy Evolution with Lenses (AGEL) survey. The authors confirm that (1) search methods using convolutional neural networks (CNNs) with visual inspection successfully identify strong gravitational lenses and (2) the lenses are at higher redshifts relative to existing surveys due to the combination of deeper and higher resolution imaging from DECam and spectroscopy that spans optical to near-infrared wavelengths \cite{c11}.

To identify lensed SNe in large imaging datasets, current surveys primarily rely on the magnification effect of gravitational lensing—searching for transients that appear brighter than expected \cite{c3} \cite{c10} \cite{c11}. These methods utilize the {\it photometry} extracted from the imaging surveys.  We propose developing an alternative method that identifies multiple images by operating directly on the pixels in the astronomical images. 
This approach is inappropriate for ground-based imaging surveys where the point spread function (PSF) is broadened by the atmosphere. The full width at half maximum (FWHM) of the PSF for the Rubin Observatory is $>0.67$ arcsec \cite{10.1093/mnras/stad664} and for DES it is $>$ 0.9 \cite{Neilsen:2016akm}, which is comparable to the typical separation of SNe lensed by galaxies \cite{c18}. However, in space-based imaging, the images are much sharper. The median PSF FWHM for HST is $\approx$ 0.067 arcsec \cite{Baltimore:STScI} and for the Roman Space Telescope it is $\approx$ 0.11 arcsec \cite{akeson2019widefieldinfraredsurvey}. Because of the better resolution of space-based imaging, it is feasible to identify multiply-imaged supernovae by the appearance of two sources on a difference image. They will be either fully resolved or appear as a blended, asymmetric PSF. 

The \textit{Roman Space Telescope} is slated to launch in 2027. % \cite{c23}.
One of the primary mission objectives for  \textit{Roman} is to investigate the nature of dark energy with a variety of methods. Observations of Type Ia supernovae (SNIa) will be one of the principal anchors of the Roman cosmology program, through traditional luminosity distance measurements. Roman will have the ability to resolve much lower image separations than ground-based SN surveys with its PSF FWHM of $\approx$ 0.11 arcseconds. In addition, {\it Roman} has a very wide field of view for a space-based observatory: spanning 0.28 square degrees, which is some 100 times larger than the Hubble Space Telescope. These characteristics make it possible for Roman imaging surveys to detect dozens of lensed SNe over a span of a few years \cite{c14}.  Roman will also detect lensed SNe at significantly higher redshifts, which lengthens the observability period due to increased time dilation and extends the cosmological leverage of the overall lensed SN sample \cite{c14}. The pixel-based CNN approach we propose in this work is designed to leverage these properties of the Roman Space Telescope and other forthcoming wide-field space-based observatories to identify lensed SNe directly from difference images.  This will facilitate rapid follow-up, because the SNe can be detected immediately--without needing a light curve, spectroscopy, or even a galaxy redshift. 

\begin{figure}[!]
\centering
\includegraphics[scale=0.28]{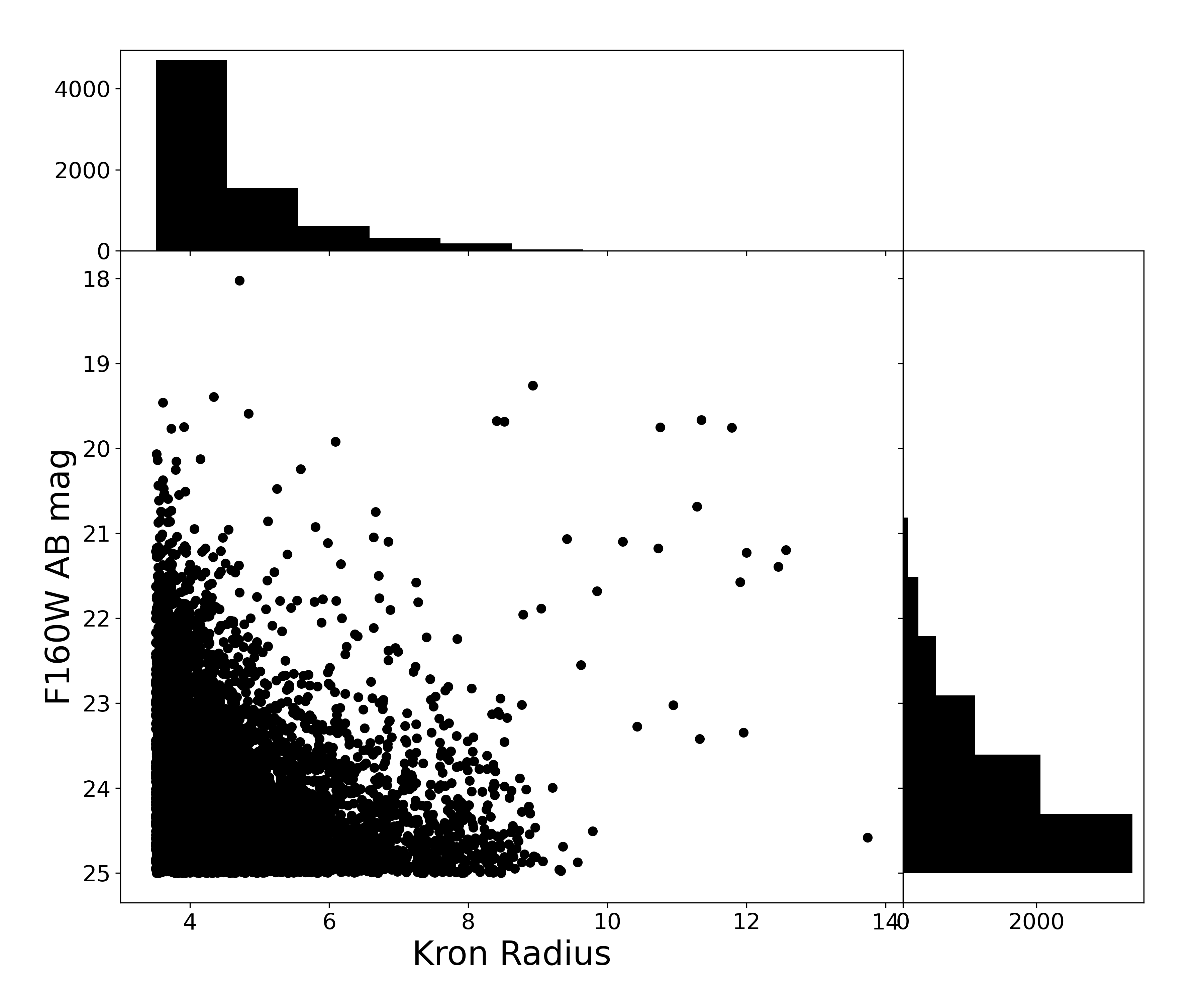}
\caption{The distribution of Kron radius and magnitudes of filter F160W with Kron radius $>$ 3.5 and magnitude between 18 and 25 for the entire available sample for filter F160W}\label{fig:overall_sample}
\end{figure}

\begin{figure}[!]
\centering
\includegraphics[scale=0.28]{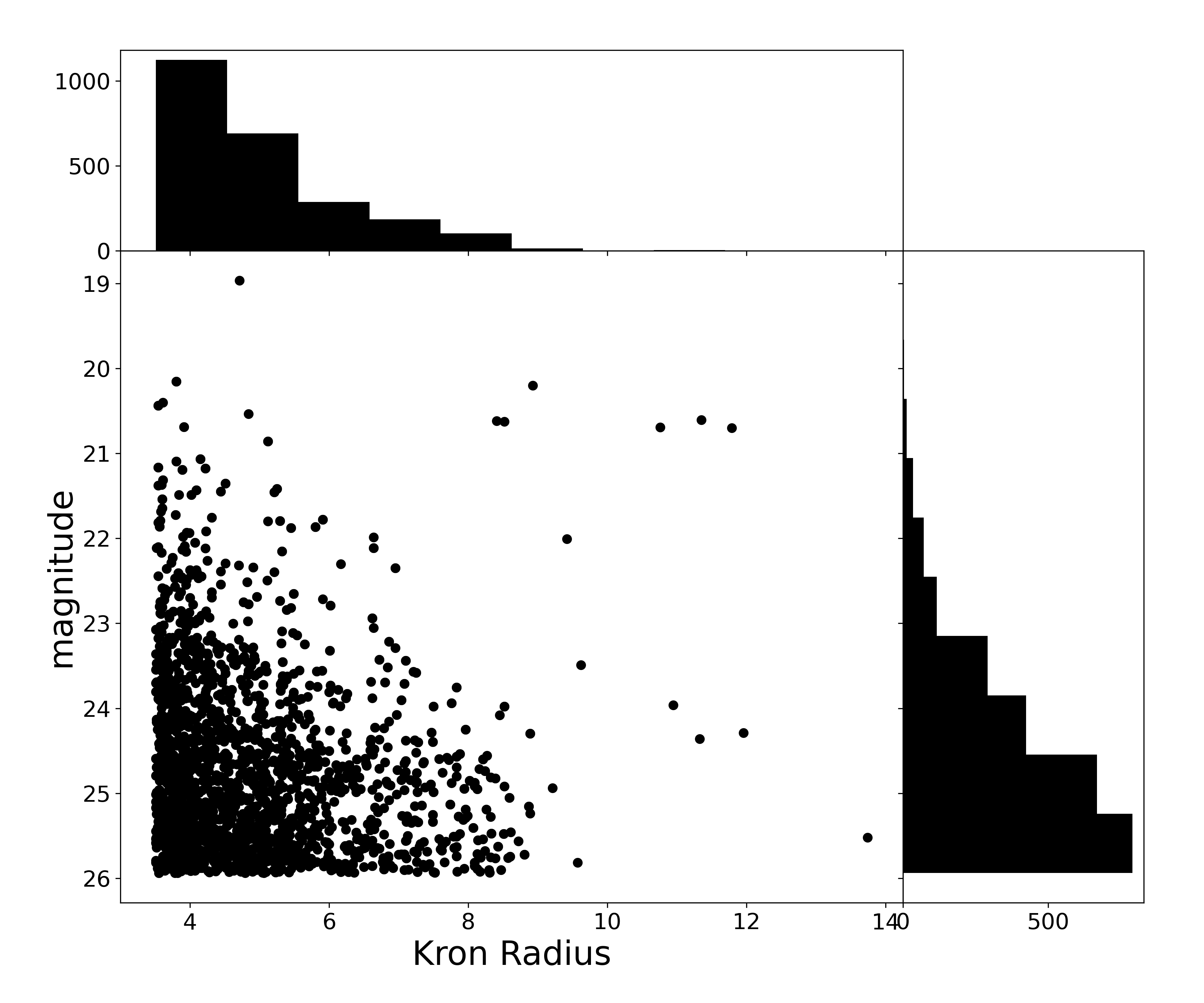}
\caption{The distribution of magnitudes and Kron radius for the overlapping region of two epochs from all filters, where we planted fakes in one epoch and took the difference from another epoch}\label{fig:filtered_sample_F160W}
\end{figure}

\begin{figure}[!]
\centering
\includegraphics[scale=0.28]{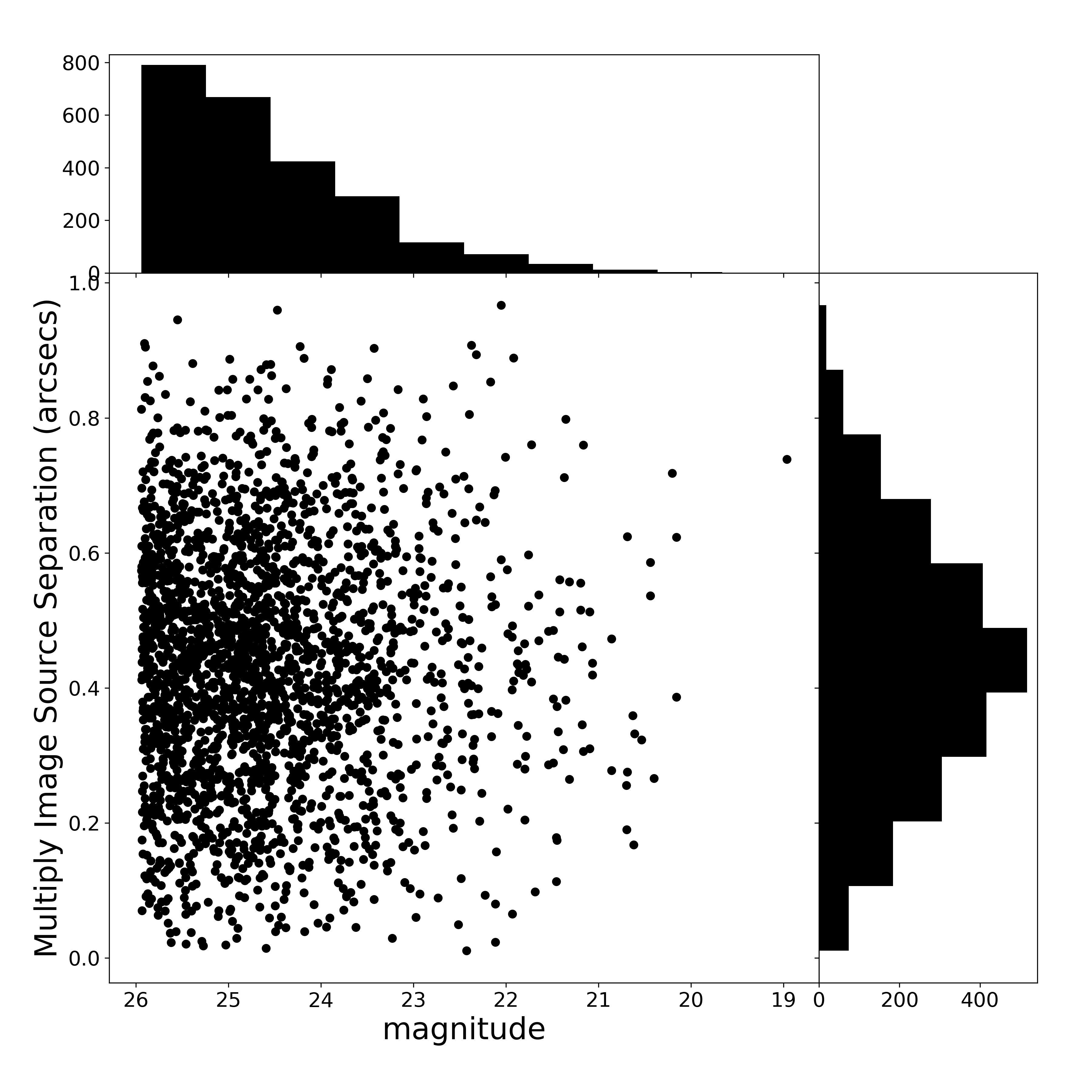}
\caption{The distribution of magnitudes and multiply-imaged source separations (in arcseconds)}\label{fig:magnitudes_vs_source_separation}
\end{figure}

\begin{figure*}
\centering
\includegraphics[scale=0.55]{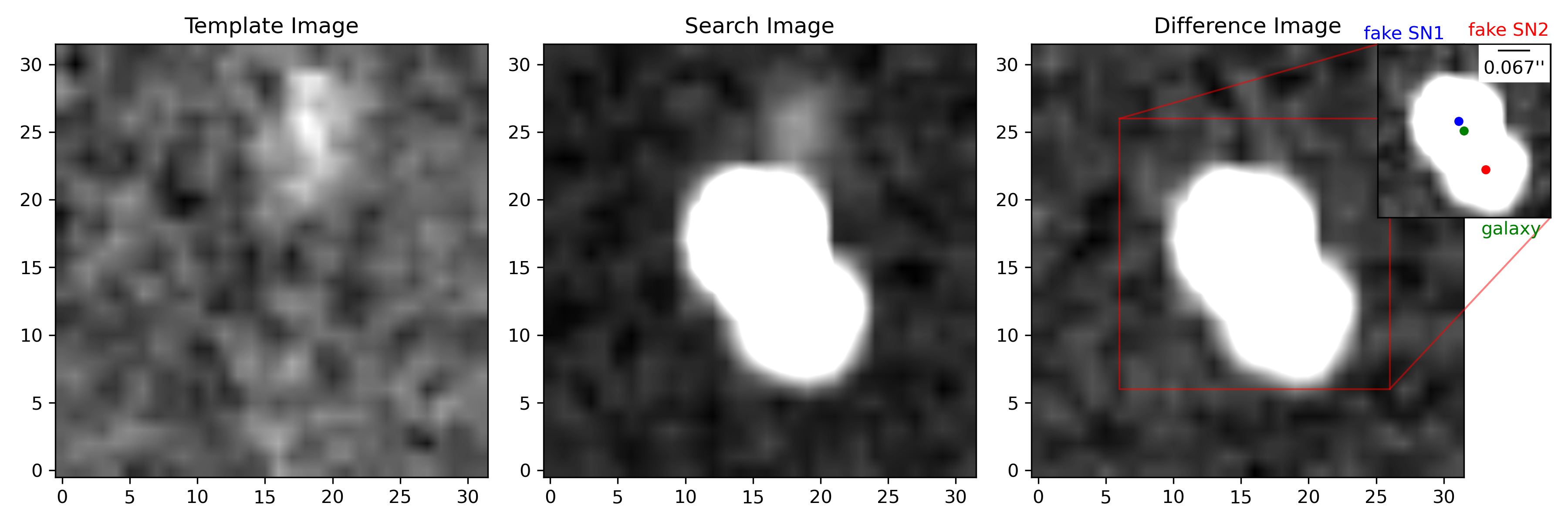}
\caption{In this triplet image, we have planted two fakes near the center galaxy, which has a Kron radius = 4.31 and a magnitude = 24.85. The image is centered on the position of the brighter of the two fake SNe. The two fake SNe are separated by 0.43 arcseconds. The fake SN1 and the center of the galaxy are separated by 0.09 arcsec. The fake SN2 and the center of the galaxy by 0.34 arcsecs. The magnitude of fake SN1 is 20.4 and the magnitude of fake SN2 is 22.9}\label{fig:two_fakes_19}
\end{figure*}

\begin{figure*}
\centering
\includegraphics[scale=0.55]{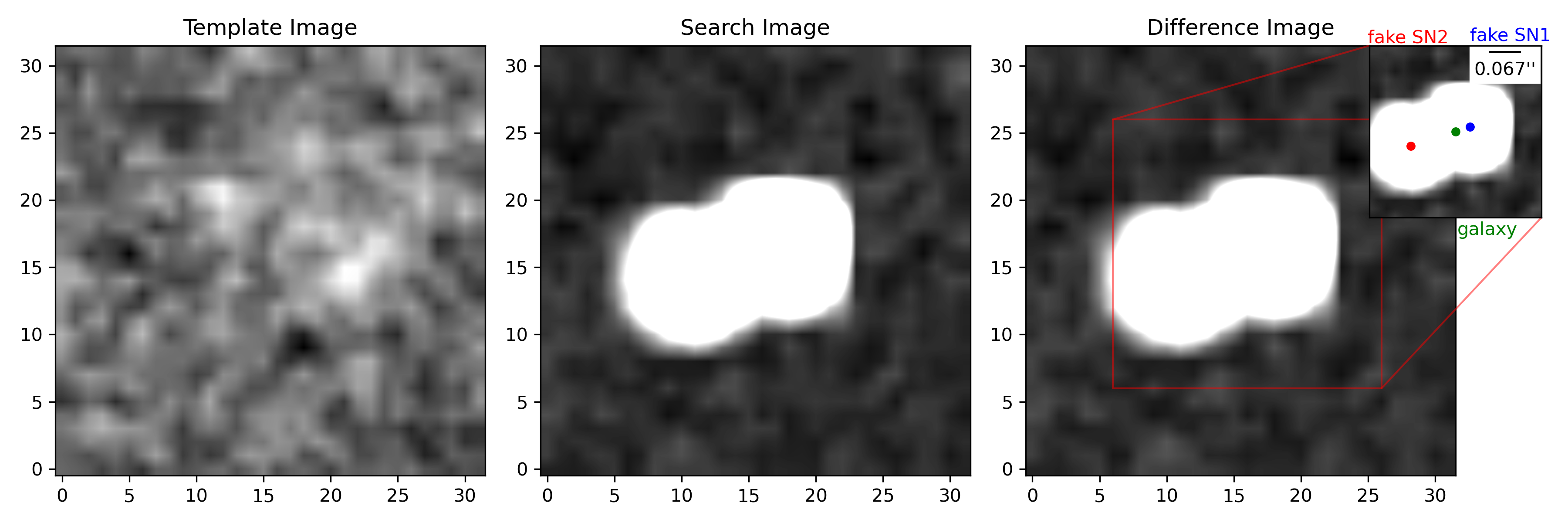}
\caption{In this triplet image, we have planted two fakes near the center galaxy, which has a Kron radius = 5.8 and a magnitude = 20.9. The image is centered on the position of the brighter of the two fake SNe. The two fake SNe are separated by 0.49 arcseconds. The fake SN1 and the center of the galaxy are separated by 0.12 arcsec. The fake SN2 and the center of the galaxy by 0.37 arcsecs. The magnitude of fake SN1 is 22.3 and the magnitude of fake SN2 is 22.8}\label{fig:two_fakes_290}
\end{figure*}

\begin{figure*}
\centering
\includegraphics[scale=0.55]{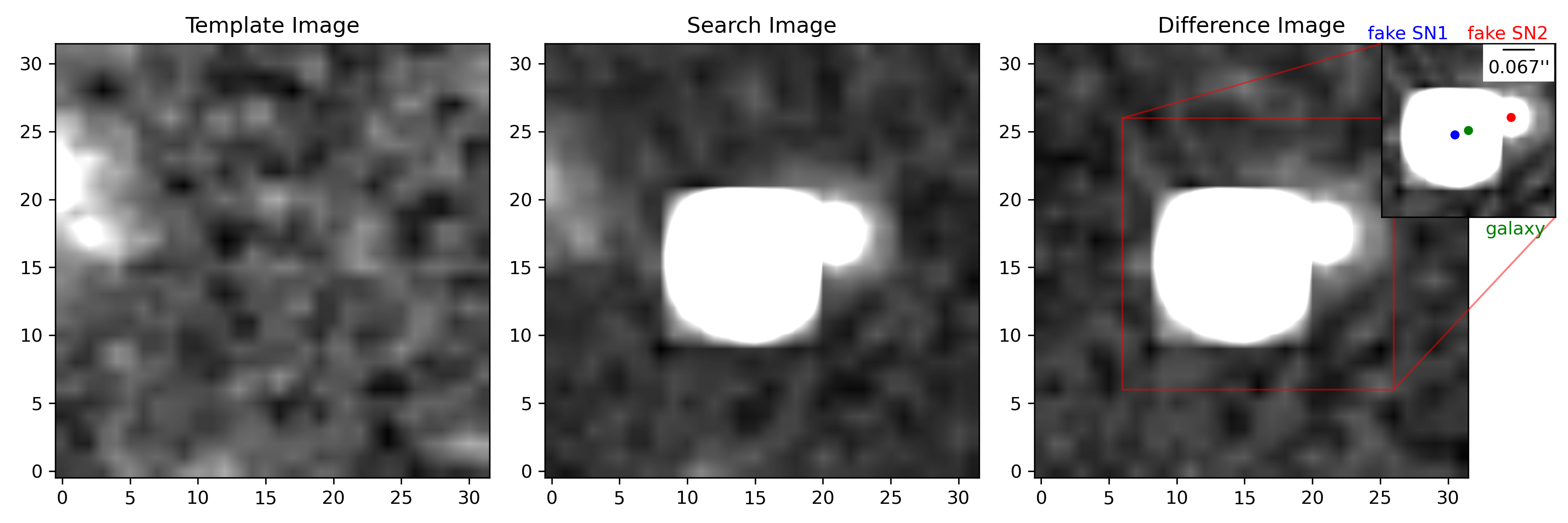}
\caption{In this triplet image, we have planted two fakes near the center galaxy, which has a Kron radius = 7.26 and a magnitude = 24.76. The image is centered on the position of the brighter of the two fake SNe. The two fake SNe are separated by 0.46 arcseconds. The fake SN1 and the center of the galaxy are separated by 0.11 arcsec. The fake SN2 and the center of the galaxy by 0.34 arcsecs. The magnitude of fake SN1 is 21.2 and the magnitude of fake SN2 is 24.6}\label{fig:single_fakes_32}
\end{figure*}

%%%%%%%%%%%%%%%%%%%%%%%%%%%%%%%%%%%%%%%%%%%%%%%%%%%%%%%%%%%%%%%%%%%%%%%%%%%%%%%%

\section{{DATA PREPARATION}}

We constructed three classes of simulated datasets of 32$\times$32-pixel images: images with zero transients, images with unlensed/single SNe, and images with multiply imaged/two-lensed GSNe. For zero-transient images, we took the difference between two consecutive epochs. For unlensed/single SNe images, we planted a fake source near a selected galaxy (with a Kron radius greater than 3.5 arcsec and an AB magnitude between 18 and 25). We took the difference between the epoch on which the fake source is planted with the previous epoch image. For multiply-imaged/two-lensed GSNe, we planted two fake sources near a selected galaxy (with a Kron radius greater than 3.5 arcsec and an AB magnitude between 18 and 25). We kept the galaxy in the middle of the two fake multiply-imaged sources. We then took the difference between the epoch at which the multiply-images sources are planted with the previous epoch image. We provided these difference images for the three classes as the input to the deep CNN for training.

As the real lensed supernovae are very rare, we constructed simulated images to build a machine learning model to identify them. We constructed our simulated training data using mosaic images from the Hubble Legacy Fields CANDELS project \cite{Grogin_2011} \cite{Koekemoer_2011}. Fig. \ref{fig:hlf_goodsn} shows the exposure time maps of the regions covered by each filter in the HLF-GOODS-N HLSP dataset. Fig. \ref{fig:hlf_goodss} shows the exposure time maps of the regions covered by each filter in the HLF-GOODS-S HLSP dataset. We begin with images from four filters that were observed in two separate observing windows (a.k.a. "epochs"): F160W (epochs 7 and 8), F775W (epochs 1 and 3), F814W (epochs 7 and 8), and F850LP (epochs 1 and 3). The original mosaic images are 25K $\times$ 25K pixels with a pixel scale of 60 milliarcseconds/pixel. They are the single-epoch mosaic images produced by the CANDELS team and released on the Hubble Legacy Archive (https://archive.stsci.edu/prepds/hlf/). We select isolated galaxies in these images to serve as proxies for the foreground lensing galaxy that would cause the appearance of a lensed SN. Massive galaxies are more likely to serve as strong gravitational lenses. Therefore, we selected galaxies for our samples by requiring a Kron radius greater than 3.5 arcsec and an AB magnitude between 18 and 25 from filter F160W (Fig. \ref{fig:overall_sample}). The Kron radius measures the size of a galaxy based on its light profile \cite{c17}. In other words, a Kron radius of 3.5 means that the radius within which 90\% of the total light of an astronomical object, such as a galaxy, is contained is 3.5 arcseconds. We also used the same galaxies (if available) in other filters to construct fake transient images. 
From the  F160W mosaic, we identify 310 galaxies that satisfy those size and magnitude requirements and also appear in the CANDELS imaging area in both epochs (Fig. \ref{fig:filtered_sample_F160W}). Fig. \ref{fig:magnitudes_vs_source_separation} shows the distribution of magnitudes and multiply-imaged source separations (in arcseconds) of these selected galaxies. Around each selected galaxy, we extracted cutout images of 32$\times$32 pixels to serve as the input images for training the neural network. 

We do not expect the CNN to actually learn to classify gravitationally lensed sources by identifying a physically accurate geometric arrangement that can plausibly be generated by a foreground lensing galaxy. Therefore, for this proof-of-concept study, we instead used a very simplified approach that merely approximates doubly-imaged SN source positioning. We planted fake sources using a PSF model near the bright galaxy at the center in one of the epochs, then subtracted the equivalent cutout from another epoch.  The result is an image triplet that includes a template image (with no fake transient in it), a search image (the image in which we planted fakes), and a difference image (search minus template).  All three of these are centered on the position of the galaxy as defined in the catalog.  This choice to center the cutout images on the galaxy reflects the aim of this study to evaluate the {\it transient classification} alone.  We are not attempting to train our network to detect the transient source. We are therefore assuming that an upstream pipeline has already detected a transient source and then sends this triplet of images to our network to ask, "Is this a lensed SN?"

We combine the three F160W cutout images as a single multi-extension fits file (MEF) with a primary header (ext=0) that includes the WCS info, then three fits extensions for each of the three images: template, search, and difference as extensions 1, 2, and 3. We repeat the same process for the other three filters (F850LP, F814W, and F775W). Figs. 
\ref{fig:two_fakes_19}, \ref{fig:two_fakes_290}, and \ref{fig:two_fakes_32} show three templates, search, and difference images we created with HLF F160W Epochs 7 and 8. In this case, we have planted a doubly-imaged SNe (two fakes).  Note that the difference image is a straightforward subtraction of the search minus the template image. That is, no convolution of the PSF is used.  This is possible for a space-based survey that has a very stable PSF. This will be true for the Roman Space Telescope as it is for the HST.

We have prepared a set of simulated gravitational lensed data for this research. The first dataset contains images of 32$\times$32 pixel size with the simulated gravitational lens at a minimum distance of 2 pixels (0.12 arcseconds) and a max distance of 10 pixels (0.6 arcseconds) from the galaxy. Also, when we feed the images to the CNN, they are already centered on the fake SN location. This is how the images would be generated by an automatic transient detection pipeline, as we expect from the Roman space telescope survey. The CNN approach we are exploring here should, therefore, be considered as a post-processing step that would be applied to all transients detected by such a pipeline. We have also built corresponding simulated images of single transients and zero transients (just the difference between two epochs). In an image with two fake sources, we planted the first fake with a flux value between 20 and 22.5 and the second with an AB magnitude value between 22.5 and 28. 

We checked how similar our simulated source configurations are to realistic lensing geometries. For this, we tested our simulated source configurations against Oguri \& Marshall \cite{c18} catalog. We found that our simulations appropriately overlap the expected range of separations by covering the most challenging cases: sources that are close to each other and close to the lens. The mean and standard deviation distance between the two sources for the Oguri \& Marshall \cite{c18} catalog was 1.50 and 0.76 arcseconds, respectively, while the mean and standard deviation distance between the two sources for our simulations are 0.44 and 0.18 arcseconds, respectively. Furthermore, the mean and standard deviation distances between the nearest source and lens for the Oguri \& Marshall \cite{c18} catalog were 0.39 and 0.25 arcseconds, respectively, while the mean and standard deviation distances between the nearest source and lens for our simulations are 0.1 and 0.04 arcseconds, respectively. That is exactly the parameter space that we would want this type of AI solution to operate in because it is the space where traditional methods of identifying single (unlensed) supernovae will be least effective.  Furthermore, when the two SN sources are very close like this, the time delays will usually be smaller, so it is more likely that both SN images will be present simultaneously (i.e., neither will have faded away) — and that simultaneity is a key assumption in this project.  

\begin{figure*}
\centering
\includegraphics[width=\textwidth]{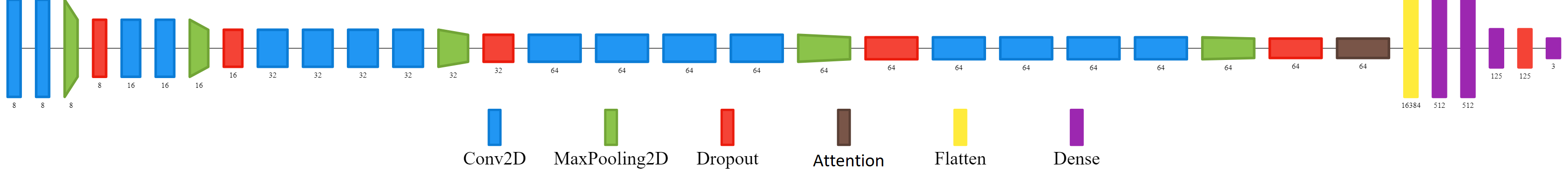}
\caption{A deep Convolutional Neural Network (CNN)  model architecture for lensed SN identification. This CNN model has 16 convolution layers, five max pool layers, one attention layer, one flatten layer, and four dense layers. This model is trained on 32$\times$32 pixel images generated from four filters from the HLF GOODS-S dataset, modified to include simulated supernovae.  The three classes are the same as defined earlier. Each class had over 2000+ images. With this model, we achieved an ROC-AUC of over 99\%. The figure is constructed using Net2Vis \cite{c2}}\label{fig:cnn}
\end{figure*}

\section{{METHODOLOGY}}

Deep Learning and Machine Learning algorithms have been extremely successful in detecting objects and patterns in large data. Machine learning models are used in all fields today; for example, in astronomy \cite{10.1093/rasti/rzaf043}, chemistry \cite{c19}, bioinformatics \cite{c20}, drug discovery \cite{c21}, and e-commerce \cite{c22}. In this work, we leverage the Convolutional Neural Network (CNN), a deep learning algorithm, to detect GSNe. VGG19 is a convolutional neural network (CNN) proposed by Karen Simonyan and Andrew Zisserman \cite{c8}. The name VGG19 originated from the Visual Geometry Group at Oxford, which developed it. It is a deep convolutional neural network with 16 CNN layers and three fully connected layers. We have built a CNN model to detect gravitational lensing inspired by the VGG19 architecture with 16 CNN layers and three fully connected layers. Our model uses CNN filters smaller than the original VGG19 model, and we also added dropout layers to prevent overfitting. The dropout layer only keeps some neurons' contributions and nullifies others' contributions. The main change we made with VGG19 is adding an attention layer \cite{c12} to the model. An attention layer introduces a mechanism within the model to enhance focus on specific parts of the input images and diminish focus on other parts. Here, the attention layer serves the purpose of aiding the network to focus on the center of the image, where the lensed SNe appear.

Fig. ~\ref{fig:cnn} shows the CNN classification model architecture similar to the VGG19 architecture to classify three classes: multiply-imaged SNe/GSNe, single source SNe/unlensed SNe and images with no SNe. To get the best results, we divided the overall dataset into two subsets: train and test, based on a 9:1 ratio. The training dataset was further divided into train and validation datasets in a 9:1 ratio for 10-fold cross-validation. We performed 10-fold cross-validation for a more robust estimate of model performance and, more importantly, to detect overfitting. As we have limited data, 10-fold cross-validation ensures that every data point is used for both training and validation. We use the RMSprop optimizer to train the model. We trained the model for a maximum of 2000 epochs with a patience value of 500, i.e., stopping model training if its performance is not improving for 500 epochs. Running the model for larger patience values did not help improve the performance of the model. We saved the model whenever its performance improved from previous epochs.

We compared our CNN results with the SExtractor's SPREAD\_MODEL \cite{c15} model. SExtractor is a program that builds a catalog of objects from an astronomical image. Although it is particularly oriented toward reducing large-scale galaxy survey data, it can perform reasonably well on moderately crowded star fields. We used the SExtractor tool to estimate the spread parameter of those sources detected in the simulation fits files through SPREAD\_MODEL. SPREAD\_MODEL is a measurement parameter in SExtractor used for star–galaxy classification. It compares how well two models fit the detected object: Point source model --- the local PSF (Point Spread Function) resampled at the object’s position, and Galaxy-like model --- a slightly “fuzzier” version of the PSF, created by convolving it with a small exponential disk (scale length = 1/16 of the PSF FWHM). The SPREAD\_MODEL is a formula that shows how the brightness of a point source spreads over a pixelated image. The SPREAD\_MODEL can help to estimate the shape and size of objects in the sky and to adjust for the effects of the air and the telescope lenses. Before running the tool, we set the parameters such as PSF FWHM, pixel scale, and magnitude zeropoint in the configuration file based on the specific HST filter type. Then, we ran the SExtractor tool in batch mode across three types of simulated image sets: zero fake, one fake, and two fakes, each generated in two different image dimensions (32 $\times$ 32 pixels). For each image, SExtractor was applied to the extension with the difference image individually. The SPREAD\_MODEL parameter and the associated error values were extracted to classify sources morphologically: values near zero indicate point sources, positive values correspond to extended sources (e.g., galaxies), and negative values typically represent detections smaller than the PSF, such as cosmic-ray hits. The SPREAD\_MODEL outputs a metric called SPREAD\_MODEL, which ranges from 0 to 1. If SPREAD\_MODEL == 0 for an object in an image, it is classified as a star. However, if SPREAD\_MODEL == 1 for an object in an image, it is classified as a galaxy. The closer the SPREAD\_MODEL is to 0 for an object, the greater the chance that that object is classified as a star. Also, the closer the SPREAD\_MODEL is to 1 for an object, the greater the chance that that object will be classified as a galaxy in an image. 

To determine how many fake supernovae (SNe) are detected in an image, first identify all SExtractor sources within 0.6 arcseconds (10 pixels) of the galaxy center. If no fake SNe were planted (Class 0), we counted all detections within this radius as false positives. For images with planted fake SNe, we matched each fake SN to the nearest SExtractor source; if the separation exceeded twice the PSF FWHM, we considered it undetected. When two fake SNe are present, we checked the brighter one first, then verified that the fainter SN is assigned to a different source to avoid duplication. The possible detection counts are 0, 1, or 2. For each detected fake SN, we recorded its SExtractor stellarity class: SPREAD\_MODEL $< 0.003$ as P (point-like) and SPREAD\_MODEL $> 0.003$ as E (extended). Finally, we determined the output Class using these rules: Class 0 for no detections, Class 1 for one detection with stellarity P, and Class 2 for either one detection with stellarity E or two detections regardless of stellarity.

\begin{table}
\begin{center}
\caption{10-fold cross-validation results on validation dataset. Here, Recall\_0 is the recall value of the class with zero fake transients, Recall\_1 is the recall value of the class with single fake transients, and Recall\_2 is the recall value of the class with two fake transients.}
 \begin{tabular}{ |l c c c c| } 
 \hline
  Filter & ROC-AUC & Recall\_0 & Recall\_1 & Recall\_2 \\
 \hline\hline
 Fold 1 & 0.9919 & 0.9641 & 0.8986 & 0.9555 \\
 Fold 2 & 0.9922 & 0.9458 & 0.8818 & 0.9381\\
 Fold 3 & 0.9949 & 0.9762 & 0.8932 & 0.9485\\
 Fold 4 & 0.9953 & 0.9955 & 0.9258 & 0.9397\\
 Fold 5 & 0.9959 & 0.9907 & 0.9113 & 0.9786\\
 Fold 6 & 0.9932 & 0.9548 & 0.8606 & 0.9546\\
 Fold 7 & 0.9969 & 0.9763 & 0.9343 & 0.9733\\
 Fold 8 & 0.9944 & 0.9957 & 0.9163 & 0.9574\\
 Fold 9 & 0.9924 & 0.9953 & 0.8853 & 0.9631\\
 Fold 10 & 0.9951 & 0.9865 & 0.8857 & 0.9676\\
 \hline
 \multicolumn{5}{|l|}{Average scores for all folds:} \\
 \multicolumn{5}{|c|}{ROC-AUC: mean = 0.9942 (std = 0.0017)} \\
 \multicolumn{5}{|c|}{Recall\_0: mean = 0.9781 (std = 0.0180)} \\
 \multicolumn{5}{|c|}{Recall\_1: mean = 0.8993 (std = 0.0226)} \\
 \multicolumn{5}{|c|}{Recall\_2: mean = 0.9576 (std = 0.0134)} \\
 \hline
\end{tabular}
\end{center}
\label{table:2}
\end{table}

\begin{table}
\begin{center}
\caption{10-fold cross-validation results on test dataset. Here, Recall\_0 is the recall value of the class with zero fake transients, Recall\_1 is the recall value of the class with single fake transients, and Recall\_2 is the recall value of the class with two fake transients.}

 \begin{tabular}{ |l c c c c| } 
 \hline
  Filter & ROC-AUC & Recall\_0 & Recall\_1 & Recall\_2 \\
 \hline\hline
 Fold 1 & 0.9925 & 0.9380 & 0.9253 & 0.9820 \\
 Fold 2 & 0.9926 & 0.9419 & 0.9046 & 0.9865\\
 Fold 3 & 0.9931 & 0.9612 & 0.9046 & 0.9685\\
 Fold 4 & 0.9926 & 0.9884 & 0.9212 & 0.9369\\
 Fold 5 & 0.9935 & 0.9612 & 0.8963 & 0.9865\\
 Fold 6 & 0.9928 & 0.9612 & 0.9087 & 0.9775\\
 Fold 7 & 0.9934 & 0.9651 & 0.8963 & 0.9775\\
 Fold 8 & 0.9930 & 0.9884 & 0.8963 & 0.9865\\
 Fold 9 & 0.9924 & 0.9767 & 0.8880 & 0.9730\\
 Fold 10 & 0.9928 & 0.9690 & 0.8714 & 0.9865\\
 \hline
 \multicolumn{5}{|l|}{Average scores for all folds:} \\
 \multicolumn{5}{|c|}{ROC-AUC: mean = 0.9919 (std = 0.0030)} \\
 \multicolumn{5}{|c|}{Recall\_0: mean = 0.9651 (std = 0.0168)} \\
 \multicolumn{5}{|c|}{Recall\_1: mean = 0.9013 (std = 0.0156)} \\
 \multicolumn{5}{|c|}{Recall\_2: mean = 0.9761 (std = 0.0152)} \\
 \hline
\end{tabular}
\end{center}
\label{table:3}
\end{table}

\section{{Results}}

We trained our deep CNN network shown in Fig. \ref{fig:cnn}, on simulated data to assess its performance on this dataset. The datasets have three classes. Class 0 and Class 1 are sets of zero transients and simulated single transients, respectively. Class 2 is a set of simulated doubly-imaged SNe. Our deep CNN model performed well on the test dataset. From Tables \ref{table:2} and \ref{table:3}, we observe that we achieved average ROC-AUC scores of 99.42\% and 99.19\% in the 10-fold cross-validation on the validation and test datasets, respectively. We achieved average recall scores of 97.81\% and 96.51\% for class 0 in the 10-fold validation and test datasets, respectively. We achieved average recall scores of 89.93\% and 90.13\% for class 1 in the 10-fold validation and test datasets, respectively. We achieved average recall scores of 95.76\% and 97.61\% for class 2 in the 10-fold validation and test datasets, respectively. Table \ref{table:hlf_goods_cnn_scores_may24} shows the results of the cross-validated fold with the highest ROC-AUC score for our model. For this fold, our model achieves an average ROC-AUC score of 99\%. This fold achieved a recall of 99\% for detecting doubly simulated transients. Our model also achieved recall scores of 90\% and 96\% for single simulated transients and zero transients, respectively. A significant limitation of this study is the reliance on synthetic data for both training and testing. While unavoidable given the absence of real Roman survey data, this introduces a risk of overly optimistic performance metrics. The models may learn features that are artifacts of the simulation rather than generalizable patterns present in real observations. Consequently, the reported ROC-AUC and recall scores might not reflect real-world performance.

\begin{table}[htp]
\begin{center}
\caption{Deep CNN classification model results for the test data fold with the highest ROC-AUC score for the cross-validated deep learning model.}\label{table:hlf_goods_cnn_scores_may24}
 \begin{tabular}{ |c c c c c| } 
 \hline
  Class & Precision & Recall & F1-score & Support \\ [0.5ex] 
 \hline\hline
 0 & 0.95 & 0.96 & 0.96 & 258 \\
 1 & 0.94 & 0.90 & 0.92 & 241 \\ 
 2 & 0.95 & 0.99 & 0.97 & 222 \\ 
   &      &      &      &       \\
 Accuracy &      &      & 0.95 & 721 \\
  \hline 
 \end{tabular}
\end{center}
\end{table}

\begin{table}[htp]
\begin{center}
\caption{SExtractor results for the test data fold with the highest ROC-AUC score for the cross-validated deep learning model.}
\label{table:hlf_goods_sextractor_scores_may24}
 \begin{tabular}{ |c c c c c| } 
 \hline
  Class & Precision & Recall & F1-score & Support \\ [0.5ex] 
 \hline\hline
 0 & 0.90 & 0.98 & 0.94 & 258 \\
 1 & 0.00 & 0.00 & 0.00 & 241 \\ 
 2 & 0.51 & 1.00 & 0.67 & 222 \\ 
   &      &      &      &       \\
 Accuracy &      &      & 0.66 & 721 \\
  \hline 
 \end{tabular}
\end{center}
 \end{table}

\begin{figure}[!]
\centering
\includegraphics[width=0.45\textwidth]{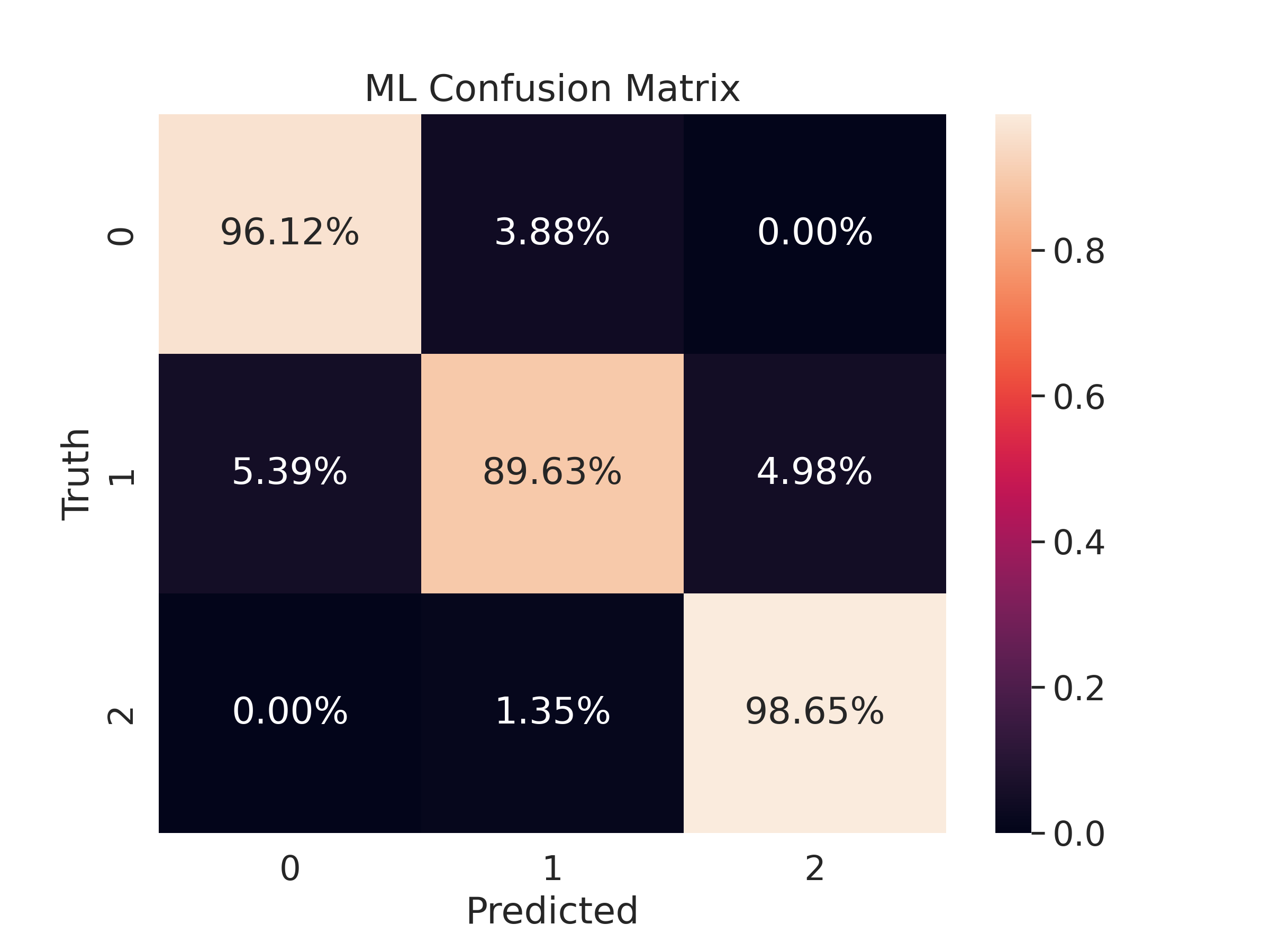}
\includegraphics[width=0.45\textwidth]{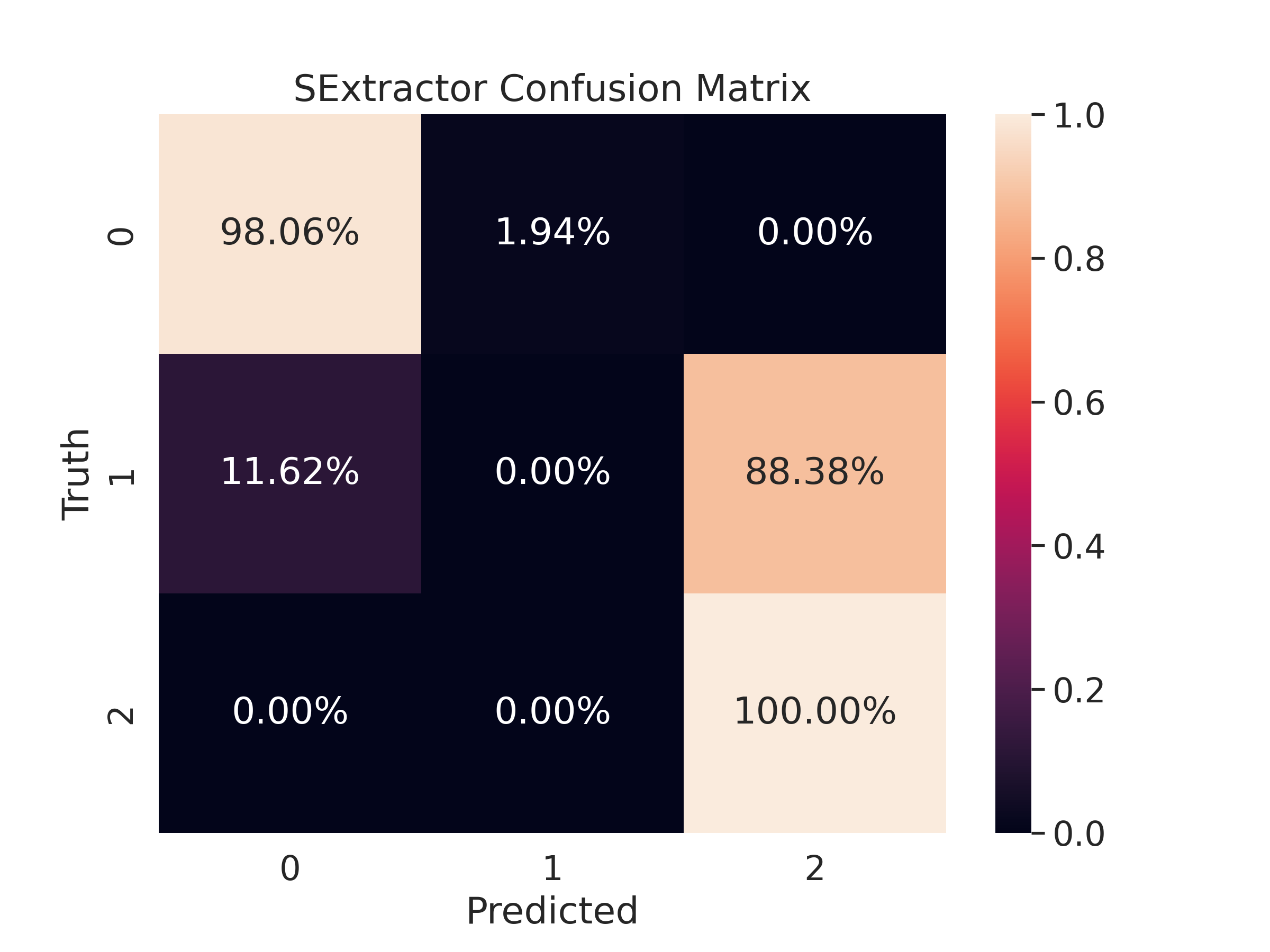}
\caption{ML and SExtractor classification scores for on test images from the fold with highest ROC-AUC score.} 
\label{fig:SExtractor_results}
\end{figure}

\subsection{{SExtractor Analysis Results}}

We also tested SExtractor's performance on this simulated dataset.  Table \ref{table:hlf_goods_sextractor_scores_may24} shows the results of the SExtractor analysis we performed on the test images. From Table \ref{table:hlf_goods_sextractor_scores_may24}, we observe that SExtractor achieved a recall score of 100\% in detecting doubly simulated transients. SExtractor also achieved recall scores of 98\% for images with zero transients. However, SExtractor cannot detect any single transient image with 0\% recall and cannot distinguish between images with a single transient and those with a doubly imaged transient.

%\clearpage

\subsection{{ML vs SExtractor}}

We compared our CNN classification results with SExtractor by testing both methods on the same test images. Fig. \ref{fig:SExtractor_results} shows the comparative results for CNN and SExtractor. We observed that both CNN and SExtractor perform very well at detecting GSNe and doubly lensed transients, with recall scores of 98.65\% and 100\%, respectively, for class 2. However, SExtractor cannot distinguish between a doubly lensed transient image and a single transient image, with a recall score of 0\% for class 1. On the other hand, our CNN model performs significantly better at distinguishing GSNe or doubly lensed transient images from single transient photos, with a recall of around 89.63\% for class 1. Both models perform very well at identifying images with no transients or zero fakes, with the CNN achieving 96.12\% recall for class 0.
In contrast, SExtractor performs slightly better, detecting over 98\% of images with zero transients. However, SExtractor also struggled to distinguish between zero transient images and single transient images, with 11.6\% of single transient images misclassified as zero transient images. In comparison, our CNN model misclassified only 5.4\% of single transient images as zero transients.

% \clearpage

\section{{Conclusions}}

Detecting multiply-imaged SNe is very important in astronomy. We have applied neural networks to detect multiply imaged SNe in the difference images from a space-based imaging survey. We constructed a simulated dataset to test the performance of neural networks in detecting multiply-imaged SNe in the difference images from a space-based imaging survey. Our simulated dataset is 32$\times$32 pixels in size. We have 2403 images in each of the three classes --- images with no transients (Class 0), images with a single transient (Class 1), and images with two transients (Class 2) — for training the CNN. We trained this simulated dataset with a deep CNN model. We achieved an overall ROC-AUC score of 99\% and a recall score of 98\% for class 2. We found that neural networks outperform a conventional algorithm (the SExtractor Star / Galaxy classification algorithm) on this task. Our deep learning model successfully distinguishes between multiply-imaged SNe and unlensed SNe.

The ability to classify gravitationally lensed supernovae from a single epoch of difference imaging is particularly valuable. Unlike methods that require multiple epochs and a light curve, early identification enables rapid follow-up observations with facilities such as JWST or HST. This is critical for obtaining multi-wavelength light curves and spectroscopy near peak brightness, which, in turn, supports more accurate classification, time-delay measurements, and determination of redshift and distance. By providing an early alert, models like the one presented here can help ensure that these rare and scientifically important events are observed in detail before they fade, maximizing the scientific return from surveys like Roman.

Future work should focus on narrowing the gap between synthetic and real survey data while maintaining the ability to classify lensed supernovae from a single-epoch difference image. This capability is essential for enabling rapid follow-up observations, so improvements should not depend on multi-epoch light curves. One option is to apply domain adaptation or transfer learning to improve model generalization from simulations to real observations. Another option is to construct hybrid datasets that combine archival survey images (e.g., HST or LSST) with simulated lenses to introduce more realistic backgrounds and noise. The synthetic difference images could be improved by including instrumental effects such as PSF variations, detector artifacts, and cosmic rays, as well as astrophysical complexities such as microlensing and irregular host-galaxy structures. Simulating Roman-specific noise characteristics and cadence irregularities would help align test conditions with what the survey will actually produce. Generative models, such as GANs or diffusion-based approaches, can also be used to generate synthetic images that better approximate real distributions. Finally, deliberately making the test set dissimilar to the training set—by varying lens mass profiles and redshift distributions, and by introducing unexpected contaminants—would provide a more rigorous assessment of model performance under realistic conditions.

\bibliography{references}

@article{c2,
  title={Net2vis--a visual grammar for automatically generating publication-tailored cnn architecture visualizations},
  author={B{\"a}uerle, Alex and Van Onzenoodt, Christian and Ropinski, Timo},
  journal={IEEE transactions on visualization and computer graphics},
  volume={27},
  number={6},
  pages={2980--2991},
  year={2021},
  publisher={IEEE}
}

@article{c3,
   title={How to Find Gravitationally Lensed Type Ia Supernovae},
   volume={834},
   ISSN={2041-8213},
   url={http://dx.doi.org/10.3847/2041-8213/834/1/L5},
   DOI={10.3847/2041-8213/834/1/l5},
   number={1},
   journal={The Astrophysical Journal Letters},
   publisher={American Astronomical Society},
   author={Goldstein, Daniel A. and Nugent, Peter E.},
   year={2016},
   month=dec, pages={L5} }

@article{c4,
  title={Multiple images of a highly magnified supernova formed by an early-type cluster galaxy lens},
  author={Kelly, Patrick L and Rodney, Steven A and Treu, Tommaso and Foley, Ryan J and Brammer, Gabriel and Schmidt, Kasper B and Zitrin, Adi and Sonnenfeld, Alessandro and Strolger, Louis-Gregory and Graur, Or and others},
  journal={Science},
  volume={347},
  number={6226},
  pages={1123--1126},
  year={2015},
  publisher={American Association for the Advancement of Science}
}

@article{c8,
  title={Very deep convolutional networks for large-scale image recognition},
  author={Simonyan, Karen and Zisserman, Andrew},
  journal={arXiv preprint arXiv:1409.1556},
  year={2014}
}

@article{c9,
  title={DeepZipper: a novel deep-learning architecture for lensed supernovae identification},
  author={Morgan, Robert and Nord, Brian and Bechtol, Keith and Gonz{\'a}lez, SJ and Buckley-Geer, E and M{\"o}ller, A and Park, JW and Kim, AG and Birrer, S and Aguena, M and others},
  journal={The Astrophysical Journal},
  volume={927},
  number={1},
  pages={109},
  year={2022},
  publisher={IOP Publishing}
}

@article{c10,
  title={DeepZipper. II. Searching for Lensed Supernovae in Dark Energy Survey Data with Deep Learning},
  author={Morgan, Robert and Nord, B and Bechtol, K and M{\"o}ller, A and Hartley, WG and Birrer, S and Gonz{\'a}lez, SJ and Martinez, M and Gruendl, RA and Buckley-Geer, EJ and others},
  journal={The Astrophysical Journal},
  volume={943},
  number={1},
  pages={19},
  year={2023},
  publisher={IOP Publishing}
}

@article{c11,
  title={Machine learning on difference image analysis: A comparison of methods for transient detection},
  author={S{\'a}nchez, Bruno and Lares, M and Beroiz, M and Cabral, JB and Gurovich, S and Qui{\~n}ones, C and Artola, R and Colazo, C and Schneiter, M and Girardini, C and others},
  journal={Astronomy and Computing},
  volume={28},
  pages={100284},
  year={2019},
  publisher={Elsevier}
}

@book{c12,
  title={Building Transformer Models with Attention: Implementing a Neural Machine Translator from Scratch in Keras},
  author={Brownlee, Jason and Cristina, Stefania and Saeed, Mehreen},
  year={2022},
  publisher={Machine Learning Mastery}
}

@mastersthesis{c14,
  title={Projected Cosmological Constraints from Strongly Lensed Supernovae with the Roman Space Telescope},
  author={Roberts-Pierel, Justin},
  year={2020},
  school={University of South Carolina}
}

@inproceedings{c15,
  title={The Dark Energy Survey data processing and calibration system},
  author={Mohr, Joseph J and Armstrong, Robert and Bertin, Emmanuel and Daues, Greg and Desai, Shantanu and Gower, Michelle and Gruendl, Robert and Hanlon, William and Kuropatkin, Nikolay and Lin, Huan and others},
  booktitle={Software and Cyberinfrastructure for Astronomy II},
  volume={8451},
  pages={121--132},
  year={2012},
  organization={SPIE}
}

@article{c16,
  title={The hubble legacy field GOODS-s photometric catalog},
  author={Whitaker, Katherine E and Ashas, Mohammad and Illingworth, Garth and Magee, Daniel and Leja, Joel and Oesch, Pascal and van Dokkum, Pieter and Mowla, Lamiya and Bouwens, Rychard and Franx, Marijn and others},
  journal={The Astrophysical Journal Supplement Series},
  volume={244},
  number={1},
  pages={16},
  year={2019},
  publisher={IOP Publishing}
}

@article{c17,
  title={Photometry of a complete sample of faint galaxies},
  author={Kron, Richard Gordon},
  journal={Astrophysical Journal Supplement Series, vol. 43, June 1980, p. 305-325. Research supported by the University of California},
  volume={43},
  pages={305--325},
  year={1980}
}

@article{c18,
  title={Gravitationally lensed quasars and supernovae in future wide-field optical imaging surveys},
  author={Oguri, Masamune and Marshall, Philip J},
  journal={Monthly Notices of the Royal Astronomical Society},
  volume={405},
  number={4},
  pages={2579--2593},
  year={2010},
  publisher={The Royal Astronomical Society}
}

@article{10.1093/mnras/stad664,
    author = {Troxel, M A and Lin, C and Park, A and Hirata, C and Mandelbaum, R and Jarvis, M and Choi, A and Givans, J and Higgins, M and Sanchez, B and Yamamoto, M and Awan, H and Chiang, J and Doré, O and Walter, C W and Zhang, T and Cohen-Tanugi, J and Gawiser, E and Hearin, A and Heitmann, K and Ishak, M and Kovacs, E and Mao, Y-Y and Wood-Vasey, M and the LSST Dark Energy Science Collaboration },
    title = {A joint Roman Space Telescope and Rubin Observatory synthetic wide-field imaging survey},
    journal = {Monthly Notices of the Royal Astronomical Society},
    volume = {522},
    number = {2},
    pages = {2801-2820},
    year = {2023},
    month = {03},
    issn = {0035-8711},
    doi = {10.1093/mnras/stad664},
    url = {https://doi.org/10.1093/mnras/stad664},
    eprint = {https://academic.oup.com/mnras/article-pdf/522/2/2801/50128648/stad664.pdf},
}

@article{Baltimore:STScI,
    author = {"Marinelli, M. and Dressel, L."},
    title = "{Wide Field Camera 3 Instrument Handbook, Version 16.0}",
    year = "2024"
}

@article{Neilsen:2016akm,
    author = "Neilsen, Eric H., Jr. and Bernstein, Gary and Gruendl, Robert and Kent, Stephen",
    title = "{Limiting Magnitude, $\tau$, $t_{eff}$, and Image Quality in DESY Year 1}",
    reportNumber = "FERMILAB-TM-2610-AE-CD",
    doi = "10.2172/1250877",
    month = "3",
    year = "2016"
}

@misc{akeson2019widefieldinfraredsurvey,
      title={The Wide Field Infrared Survey Telescope: 100 Hubbles for the 2020s}, 
      author={Rachel Akeson and Lee Armus and Etienne Bachelet and Vanessa Bailey and Lisa Bartusek and Andrea Bellini and Dominic Benford and David Bennett and Aparna Bhattacharya and Ralph Bohlin and Martha Boyer and Valerio Bozza and Geoffrey Bryden and Sebastiano Calchi Novati and Kenneth Carpenter and Stefano Casertano and Ami Choi and David Content and Pratika Dayal and Alan Dressler and Olivier Doré and S. Michael Fall and Xiaohui Fan and Xiao Fang and Alexei Filippenko and Steven Finkelstein and Ryan Foley and Steven Furlanetto and Jason Kalirai and B. Scott Gaudi and Karoline Gilbert and Julien Girard and Kevin Grady and Jenny Greene and Puragra Guhathakurta and Chen Heinrich and Shoubaneh Hemmati and David Hendel and Calen Henderson and Thomas Henning and Christopher Hirata and Shirley Ho and Eric Huff and Anne Hutter and Rolf Jansen and Saurabh Jha and Samson Johnson and David Jones and Jeremy Kasdin and Patrick Kelly and Robert Kirshner and Anton Koekemoer and Jeffrey Kruk and Nikole Lewis and Bruce Macintosh and Piero Madau and Sangeeta Malhotra and Kaisey Mandel and Elena Massara and Daniel Masters and Julie McEnery and Kristen McQuinn and Peter Melchior and Mark Melton and Bertrand Mennesson and Molly Peeples and Matthew Penny and Saul Perlmutter and Alice Pisani and Andrés Plazas and Radek Poleski and Marc Postman and Clément Ranc and Bernard Rauscher and Armin Rest and Aki Roberge and Brant Robertson and Steven Rodney and James Rhoads and Jason Rhodes and Russell Ryan Jr. and Kailash Sahu and David Sand and Dan Scolnic and Anil Seth and Yossi Shvartzvald and Karelle Siellez and Arfon Smith and David Spergel and Keivan Stassun and Rachel Street and Louis-Gregory Strolger and Alexander Szalay and John Trauger and M. A. Troxel and Margaret Turnbull and Roeland van der Marel and Anja von der Linden and Yun Wang and David Weinberg and Benjamin Williams and Rogier Windhorst and Edward Wollack and Hao-Yi Wu and Jennifer Yee and Neil Zimmerman},
      year={2019},
      eprint={1902.05569},
      archivePrefix={arXiv},
      primaryClass={astro-ph.IM},
      url={https://arxiv.org/abs/1902.05569}, 
}

@article{10.1093/rasti/rzaf043,
    author = {Kirmani, Fawad and Unni, Ananthavishnu S and Kulkarni, Varsha P and Lackey, Kyle and Rose, John R},
    title = {Detecting polar ring galaxies via deep learning},
    journal = {RAS Techniques and Instruments},
    volume = {4},
    pages = {rzaf043},
    year = {2025},
    month = {10},
    issn = {2752-8200},
    doi = {10.1093/rasti/rzaf043},
    url = {https://doi.org/10.1093/rasti/rzaf043},
    eprint = {https://academic.oup.com/rasti/article-pdf/doi/10.1093/rasti/rzaf043/64484989/rzaf043.pdf},
}

@article{c19,
  title={Combining machine learning and computational chemistry for predictive insights into chemical systems},
  author={Keith, John A and Vassilev-Galindo, Valentin and Cheng, Bingqing and Chmiela, Stefan and Gastegger, Michael and Muller, Klaus-Robert and Tkatchenko, Alexandre},
  journal={Chemical reviews},
  volume={121},
  number={16},
  pages={9816--9872},
  year={2021},
  publisher={ACS Publications}
}

@inproceedings{c20,
  title={Exploring Machine Learning Techniques To Improve Peptide Identification},
  author={Kirmani, Fawad and Lane, Bryan Jeremy and Rose, John R},
  booktitle={2019 IEEE 19th International Conference on Bioinformatics and Bioengineering (BIBE)},
  pages={66--71},
  year={2019},
  organization={IEEE}
}

@article{c21,
  title={Machine learning in drug discovery: a review},
  author={Dara, Suresh and Dhamercherla, Swetha and Jadav, Surender Singh and Babu, CH Madhu and Ahsan, Mohamed Jawed},
  journal={Artificial intelligence review},
  volume={55},
  number={3},
  pages={1947--1999},
  year={2022},
  publisher={Springer}
}

@article{c22,
  title={Exposing, formalizing and reasoning over the latent semantics of tags in multimodal data sources},
  author={Tyler, John and Pastor, Jon and Huhns, Michael N and Kirmani, Shad and Du, Hongying},
  journal={Applied Ontology},
  volume={8},
  number={2},
  pages={95--130},
  year={2013},
  publisher={IOS Press}
}

@article{Koekemoer_2011,
   title={CANDELS: THE COSMIC ASSEMBLY NEAR-INFRARED DEEP EXTRAGALACTIC LEGACY SURVEY—THE
                    HUBBLE SPACE TELESCOPE
                    OBSERVATIONS, IMAGING DATA PRODUCTS, AND MOSAICS},
   volume={197},
   ISSN={1538-4365},
   url={http://dx.doi.org/10.1088/0067-0049/197/2/36},
   DOI={10.1088/0067-0049/197/2/36},
   number={2},
   journal={The Astrophysical Journal Supplement Series},
   publisher={American Astronomical Society},
   author={Koekemoer, Anton M. and Faber, S. M. and Ferguson, Henry C. and Grogin, Norman A. and Kocevski, Dale D. and Koo, David C. and Lai, Kamson and Lotz, Jennifer M. and Lucas, Ray A. and McGrath, Elizabeth J. and Ogaz, Sara and Rajan, Abhijith and Riess, Adam G. and Rodney, Steve A. and Strolger, Louis and Casertano, Stefano and Castellano, Marco and Dahlen, Tomas and Dickinson, Mark and Dolch, Timothy and Fontana, Adriano and Giavalisco, Mauro and Grazian, Andrea and Guo, Yicheng and Hathi, Nimish P. and Huang, Kuang-Han and van der Wel, Arjen and Yan, Hao-Jing and Acquaviva, Viviana and Alexander, David M. and Almaini, Omar and Ashby, Matthew L. N. and Barden, Marco and Bell, Eric F. and Bournaud, Frédéric and Brown, Thomas M. and Caputi, Karina I. and Cassata, Paolo and Challis, Peter J. and Chary, Ranga-Ram and Cheung, Edmond and Cirasuolo, Michele and Conselice, Christopher J. and Cooray, Asantha Roshan and Croton, Darren J. and Daddi, Emanuele and Davé, Romeel and de Mello, Duilia F. and de Ravel, Loic and Dekel, Avishai and Donley, Jennifer L. and Dunlop, James S. and Dutton, Aaron A. and Elbaz, David and Fazio, Giovanni G. and Filippenko, Alexei V. and Finkelstein, Steven L. and Frazer, Chris and Gardner, Jonathan P. and Garnavich, Peter M. and Gawiser, Eric and Gruetzbauch, Ruth and Hartley, Will G. and Häussler, Boris and Herrington, Jessica and Hopkins, Philip F. and Huang, Jia-Sheng and Jha, Saurabh W. and Johnson, Andrew and Kartaltepe, Jeyhan S. and Khostovan, Ali A. and Kirshner, Robert P. and Lani, Caterina and Lee, Kyoung-Soo and Li, Weidong and Madau, Piero and McCarthy, Patrick J. and McIntosh, Daniel H. and McLure, Ross J. and McPartland, Conor and Mobasher, Bahram and Moreira, Heidi and Mortlock, Alice and Moustakas, Leonidas A. and Mozena, Mark and Nandra, Kirpal and Newman, Jeffrey A. and Nielsen, Jennifer L. and Niemi, Sami and Noeske, Kai G. and Papovich, Casey J. and Pentericci, Laura and Pope, Alexandra and Primack, Joel R. and Ravindranath, Swara and Reddy, Naveen A. and Renzini, Alvio and Rix, Hans-Walter and Robaina, Aday R. and Rosario, David J. and Rosati, Piero and Salimbeni, Sara and Scarlata, Claudia and Siana, Brian and Simard, Luc and Smidt, Joseph and Snyder, Diana and Somerville, Rachel S. and Spinrad, Hyron and Straughn, Amber N. and Telford, Olivia and Teplitz, Harry I. and Trump, Jonathan R. and Vargas, Carlos and Villforth, Carolin and Wagner, Cory R. and Wandro, Pat and Wechsler, Risa H. and Weiner, Benjamin J. and Wiklind, Tommy and Wild, Vivienne and Wilson, Grant and Wuyts, Stijn and Yun, Min S.},
   year={2011},
   month=dec, pages={36} }

@article{Grogin_2011,
   title={CANDELS: THE COSMIC ASSEMBLY NEAR-INFRARED DEEP EXTRAGALACTIC LEGACY SURVEY},
   volume={197},
   ISSN={1538-4365},
   url={http://dx.doi.org/10.1088/0067-0049/197/2/35},
   DOI={10.1088/0067-0049/197/2/35},
   number={2},
   journal={The Astrophysical Journal Supplement Series},
   publisher={American Astronomical Society},
   author={Grogin, Norman A. and Kocevski, Dale D. and Faber, S. M. and Ferguson, Henry C. and Koekemoer, Anton M. and Riess, Adam G. and Acquaviva, Viviana and Alexander, David M. and Almaini, Omar and Ashby, Matthew L. N. and Barden, Marco and Bell, Eric F. and Bournaud, Frédéric and Brown, Thomas M. and Caputi, Karina I. and Casertano, Stefano and Cassata, Paolo and Castellano, Marco and Challis, Peter and Chary, Ranga-Ram and Cheung, Edmond and Cirasuolo, Michele and Conselice, Christopher J. and Cooray, Asantha Roshan and Croton, Darren J. and Daddi, Emanuele and Dahlen, Tomas and Davé, Romeel and de Mello, Duília F. and Dekel, Avishai and Dickinson, Mark and Dolch, Timothy and Donley, Jennifer L. and Dunlop, James S. and Dutton, Aaron A. and Elbaz, David and Fazio, Giovanni G. and Filippenko, Alexei V. and Finkelstein, Steven L. and Fontana, Adriano and Gardner, Jonathan P. and Garnavich, Peter M. and Gawiser, Eric and Giavalisco, Mauro and Grazian, Andrea and Guo, Yicheng and Hathi, Nimish P. and Häussler, Boris and Hopkins, Philip F. and Huang, Jia-Sheng and Huang, Kuang-Han and Jha, Saurabh W. and Kartaltepe, Jeyhan S. and Kirshner, Robert P. and Koo, David C. and Lai, Kamson and Lee, Kyoung-Soo and Li, Weidong and Lotz, Jennifer M. and Lucas, Ray A. and Madau, Piero and McCarthy, Patrick J. and McGrath, Elizabeth J. and McIntosh, Daniel H. and McLure, Ross J. and Mobasher, Bahram and Moustakas, Leonidas A. and Mozena, Mark and Nandra, Kirpal and Newman, Jeffrey A. and Niemi, Sami-Matias and Noeske, Kai G. and Papovich, Casey J. and Pentericci, Laura and Pope, Alexandra and Primack, Joel R. and Rajan, Abhijith and Ravindranath, Swara and Reddy, Naveen A. and Renzini, Alvio and Rix, Hans-Walter and Robaina, Aday R. and Rodney, Steven A. and Rosario, David J. and Rosati, Piero and Salimbeni, Sara and Scarlata, Claudia and Siana, Brian and Simard, Luc and Smidt, Joseph and Somerville, Rachel S. and Spinrad, Hyron and Straughn, Amber N. and Strolger, Louis-Gregory and Telford, Olivia and Teplitz, Harry I. and Trump, Jonathan R. and van der Wel, Arjen and Villforth, Carolin and Wechsler, Risa H. and Weiner, Benjamin J. and Wiklind, Tommy and Wild, Vivienne and Wilson, Grant and Wuyts, Stijn and Yan, Hao-Jing and Yun, Min S.},
   year={2011},
   month=dec, pages={35} }

\end{document}